\documentclass[twocolumn,secnumarabic,amssymb, nobibnotes, aps, prd]{revtex4}
\usepackage{graphicx}
\usepackage{dcolumn}
\usepackage{bm}
\newcommand{\bfx}{\mbox{\boldmath$x$}}
\newcommand{\bfv}{\mbox{\boldmath$v$}}
\newcommand{\re}{r_e}
\newcommand{\tff}{T_{\rm ff}}
\newcommand{\trelax}{T_{\rm relax}}
\newcommand{\Sbg}{S_{\rm\scriptscriptstyle BG}}
\newcommand{\Sts}{S_q}
\newcommand{\Dcrit}{D_{\rm crit}}
\begin{document}
%
%
%
%
%
\title{Long-term Evolution of Stellar Self-Gravitating System away from 
the Thermal Equilibrium : connection with non-extensive statistics}
%
%
%
\author{Atsushi Taruya}%
\affiliation{Research Center for the Early Universe, School of Science, 
	University of Tokyo, Tokyo 113-0033, Japan}
\author{Masa-aki Sakagami}%
\affiliation{Department of Fundamental Sciences, FIHS, Kyoto University, 
Kyoto 606-8501, Japan}
\date{\today}%
%
%
%
%
%
\begin{abstract}
With particular attention to the recently postulated introduction of a 
non-extensive generalization of Boltzmann-Gibbs statistics,  
we study the long-term stellar dynamical evolution of self-gravitating systems  
on timescales much longer than the two-body relaxation time. In a self-gravitating 
$N$-body system confined in an adiabatic wall, we show that the 
quasi-equilibrium sequence arising from the Tsallis entropy, so-called 
{\it stellar polytropes}, plays an important role in characterizing the 
transient states away from the Boltzmann-Gibbs equilibrium state. 
\end{abstract}
%
%
%
\pacs{98.10.+z, 04.40.-b, 05.70.-a, 64.60.Fr}
\maketitle
%
%
%
%
%
%
%
%
%
%
%
%
%
%
%
%
%
{\it Introduction.---} 
The long-term evolution of self-gravitating many-body system 
is an old problem with a rich history in astrophysics. 
The problem, in nature, involves the long-range nature of 
attractive gravity and is fundamentally connected with 
statistical mechanics and thermodynamics. 
Historically, the important consequence from the thermodynamical  
arguments has arisen in 1960s, known as the 
{\it gravothermal catastrophe}, i.e., thermodynamic instability due to 
the negative specific heat\cite{BT_S}. Originally, the gravothermal 
catastrophe has been investigated in a very idealized situation, i.e., 
a stellar system 
confined in a spherical cavity\cite{A_LW}. Owing to the maximum 
entropy principle, the existence of an unstable thermal state has been found from 
the standard analysis of statistical mechanics with a particular attention  
to the Boltzmann-Gibbs(BG) entropy: 
\begin{equation}
\Sbg=-\int d^3\bfx d^3\bfv\,f(\bfx,\bfv) \ln f(\bfx, \bfv), 
\label{eq: S_BG}
\end{equation}
where $f(\bfx,\bfv)$ denotes the one-particle distribution function 
defined in  phase-space $(\bfx,\bfv)$.

Since 1960s, the standard approach using the BG entropy 
has dramatically improved our view of the late-time phase of 
the globular cluster as a real astronomical system\cite{MH}, 
however, from a thermodynamic point of view, 
peculiarities of the thermal property in self-gravitating systems such 
as negative specific heat, as well as the non-equilibrium 
properties away from the BG state have not yet been understood completely.

In this Letter, aiming at a better understanding of the (non-equilibrium) 
thermodynamic properties of stellar self-gravitating systems, 
we present a set of long-term $N$-body simulations, the timescale of 
which is much longer than the relaxation time.  
With a particular emphasis to the recent application of the non-extensive 
generalization of BG statistics, we focus on the stellar dynamical 
evolution in an isolated star cluster before self-similar core-collapse\cite{C1980}. 
We show that the quasi-equilibrium sequence arising 
from the Tsallis entropy\cite{T1988} plays an important role in 
characterizing the non-equilibrium evolution of a self-gravitating system.   
%
%
%
%
%
%
%
%
%
%
%
%
%
%
%
%
%
%
%
%
%
%
%
%
%
%
%
%
%

{\it $N$-body simulations.---}
The $N$-body experiment considered here is the same 
situation as investigated in classic papers (\cite{A_LW}, see 
also Ref.\cite{EFM1997}). 
That is, we confine the 
$N$ particles interacted via Newton gravity in a spherical adiabatic wall, 
which reverses the radial components of the velocity if the particle reaches 
the wall. 
Without loss of generality, we set the units as $G=M=\re=1$, where 
$G$ is gravitational constant, $M$ and $\re$ is the total mass of the system 
and the radius of the adiabatic wall, respectively.   
Note that the typical timescales appearing in this system are 
the free-fall time, $\tff=(G\rho)^{-1/2}$, and the global relaxation time 
driven by the two-body encounter, $\trelax=(0.1N/\ln N) \tff$ \cite{BT_S}, 
which are basically scaled as $\tff\sim1$ and 
$\trelax\sim 0.1N/\ln N$ in our units. 
%
%
%
%
%
%
%
%
%
%
%
%
To perform an expensive N-body calculation,  
we used a special-purpose hardware, GRAPE-6, 
which is especially designed to accelerate the gravitational 
force calculations for collisional $N$-body systems\cite{M2002}. 
With this implementation, the $4$th-order 
Hermite integrator with individual time step\cite{MA1992} 
can work efficiently, which is suited for probing the relaxation 
process in denser core regions with an appropriate accuracy. 
We adopt the Plummer softened potential, $\phi=1/\sqrt{r^2+\epsilon^2}$ 
with a softening length $\epsilon$ of $1/512$ and $1/2048$. 
%
%
%
%
%
%
%
%
%
%
%
%
%
%
%
%
%
%
%
%
%
\begin{figure}[ht]
\includegraphics[width=6.8cm]{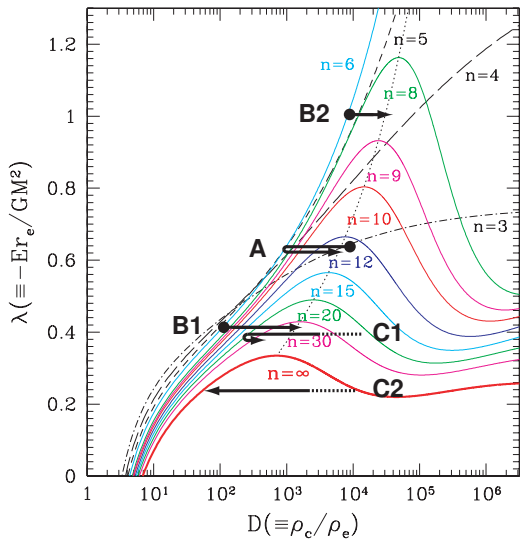}
\caption{\label{fig:lambda} Equilibrium sequences of stellar polytrope 
and isothermal distribution ($n=\infty$) 
in the energy-density contrast relation, $\lambda\equiv-r_eE/(GM^2)$ v.s. 
$D\equiv\rho_c/\rho_e$. The thick arrows denote the evolutionary tracks in 
each simulation run. }
\end{figure}
%
%
%
%
%
%
%
%
\begin{table*}
\caption{\label{tab:initial_model}Initial distributions and their 
evolutionary states}
\begin{ruledtabular}
\begin{tabular}{cclccc}
 run $\#$ & initial distribution & parameters & $\#$ of particles  & transient state & final state \\
\hline
 A & stellar polytrope($n=3$) & $D=10,000$  & 2,048 &stellar polytrope& collapse \\
 B1& stellar polytrope($n=6$) & $D=110$     & 2,048 &stellar polytrope& collapse \\ 
 B2& stellar polytrope($n=6$) & $D=10,000$  & 2,048 &stellar polytrope& collapse \\
 C1& Hernquist model          & $a/r_e=0.5$ & 8,192 &stellar polytrope& collapse \\
 C2& Hernquist model          & $a/r_e=1.0$ & 8,192 & none &  isothermal \\
\end{tabular}
\end{ruledtabular}
\end{table*}

In the present numerical simulation, 
the choice of the initial condition is an important step. 
Here, we set the initial distribution 
to the stationary state of Poisson-Vlassov equation, 
i.e., dynamical equilibrium for a spherical system with isotropic velocity 
distribution. According to the 
Jeans theorem\cite{BT_S}, the one-particle distribution function 
$f(\bfx,\bfv)$ can be expressed as a function of specific energy, 
$\varepsilon=v^2/2+\Phi(r)$ with $r$ and $\Phi$ being the radius and the 
gravitational potential. 
Then keeping the energy and the mass constant, 
the thermal equilibrium of ordinary extensive statistics 
derived from the maximum entropy principle 
of the BG entropy (\ref{eq: S_BG}) reduces to the exponential distribution, 
so-called {\it isothermal} distribution given by 
$f(\varepsilon)\propto e^{-\beta\varepsilon}$, 
which effectively satisfies the equation of state, 
$P(r)\propto \rho(r)$, where $P(r)$ is pressure and $\rho(r)$ is mass 
density\cite{BT_S,A_LW}.

On the other hand, as another possibility, 
one considers the extremum state of Tsallis' non-extensive 
entropy \cite{T1988}: 
\begin{equation}
\Sts=-\int d^3\bfx d^3\bfv\,[\{f(\bfx,\bfv)\}^q-f(\bfx,\bfv)]/(1-q), 
\label{eq: S_TS}
\end{equation}
which might be of particular importance in describing the quasi-equilibrium 
state away from the BG state\cite{T_AO}. In this case, 
the maximum entropy principle leads to the power-law distribution, 
$f(\varepsilon)\propto (\Phi_0-\varepsilon)^{1/(q-1)}$
\cite{PP1993,TS2002a,TS2003b,TS2002c}, referred to as the 
{\it stellar polytrope}\cite{BT_S}. It satisfies 
the polytropic equation of state, $P(r)\propto\rho(r)^{1+1/n}$,  
and the polytrope index $n$ is related to the $q$-parameter as 
$n=1/(q-1)+3/2$\cite{notice_1}. 
Provided the polytrope index $n$, the equilibrium 
structure can be determined by solving the L\'ane-Emden equation
\cite{C1939} and using this solution, the relationship between 
the dimensionless energy $\lambda\equiv-\re E/(GM^2)$ and 
the density contrast $D\equiv\rho_c/\rho_e$, the core density divided by 
the edge density, 
can be drawn (see Fig.\ref{fig:lambda}, \cite{notice_2}). Note that the limit 
$n\to\infty$ (or $q\to1$) corresponds to the isothermal distribution 
derived from the BG entropy (\ref{eq: S_BG}).

\begin{figure*}[htb]
\includegraphics[height=4.4cm]{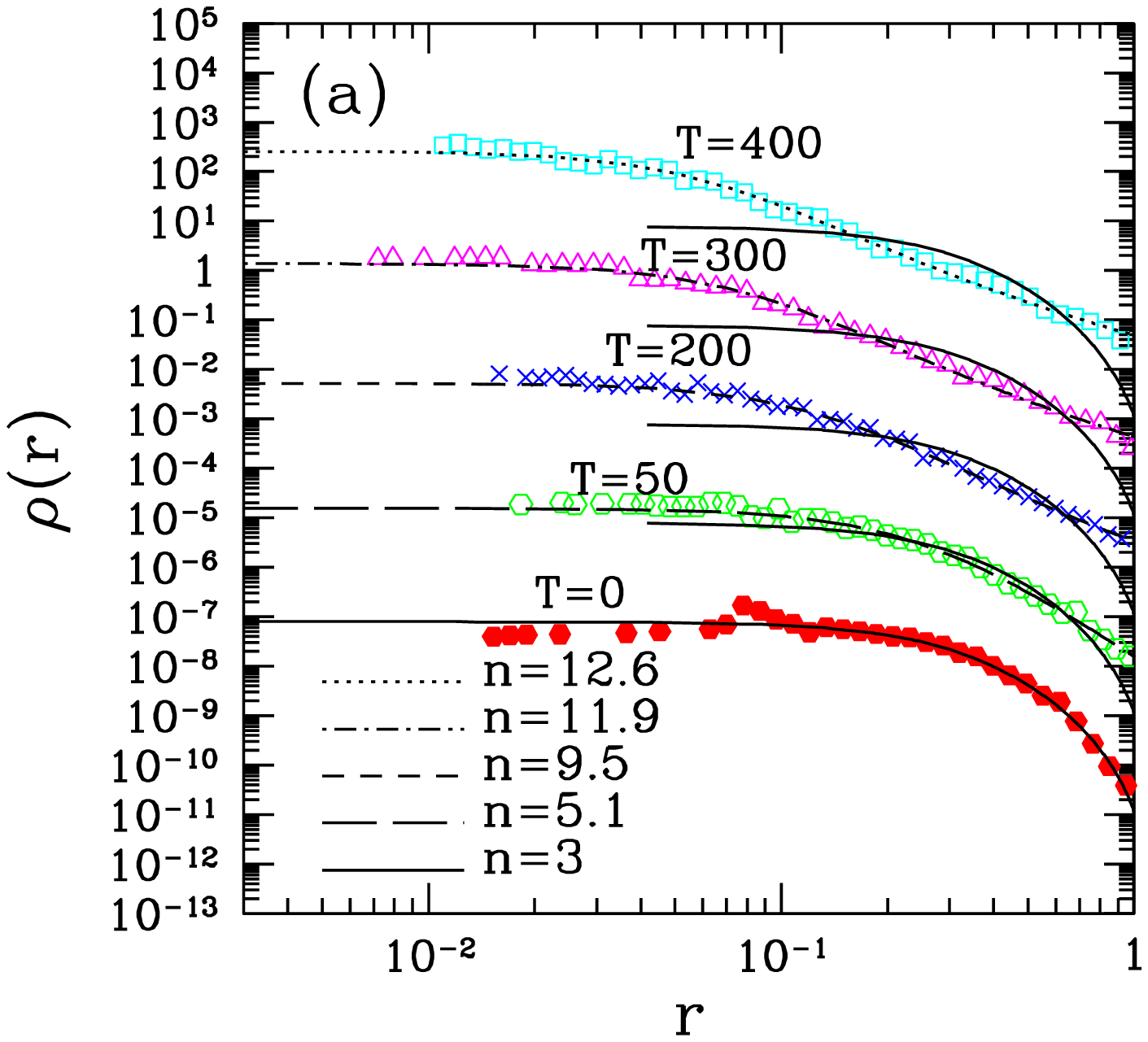}
\hspace*{0.4cm}
\includegraphics[height=4.4cm]{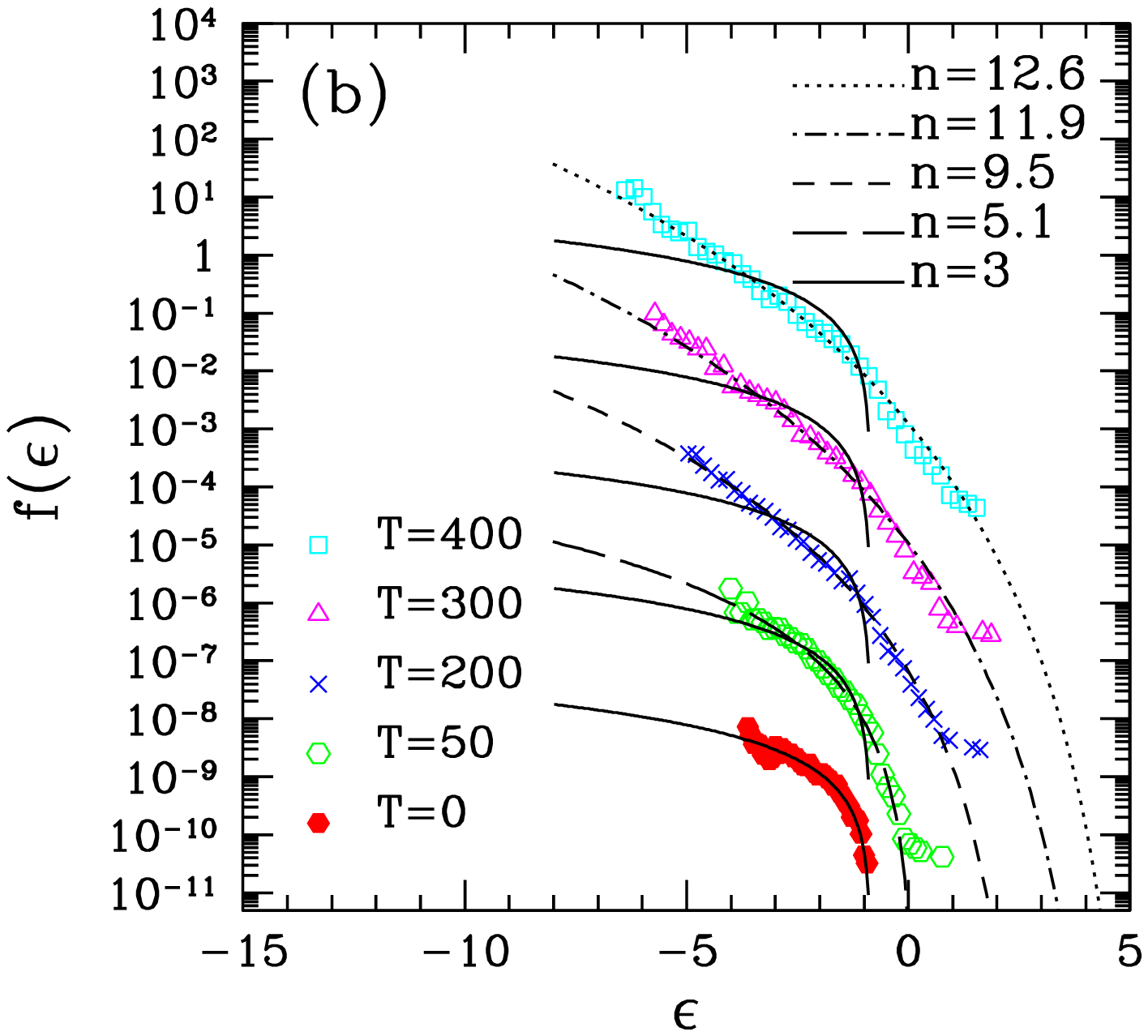}
\hspace*{0.4cm}
\includegraphics[width=4.6cm]{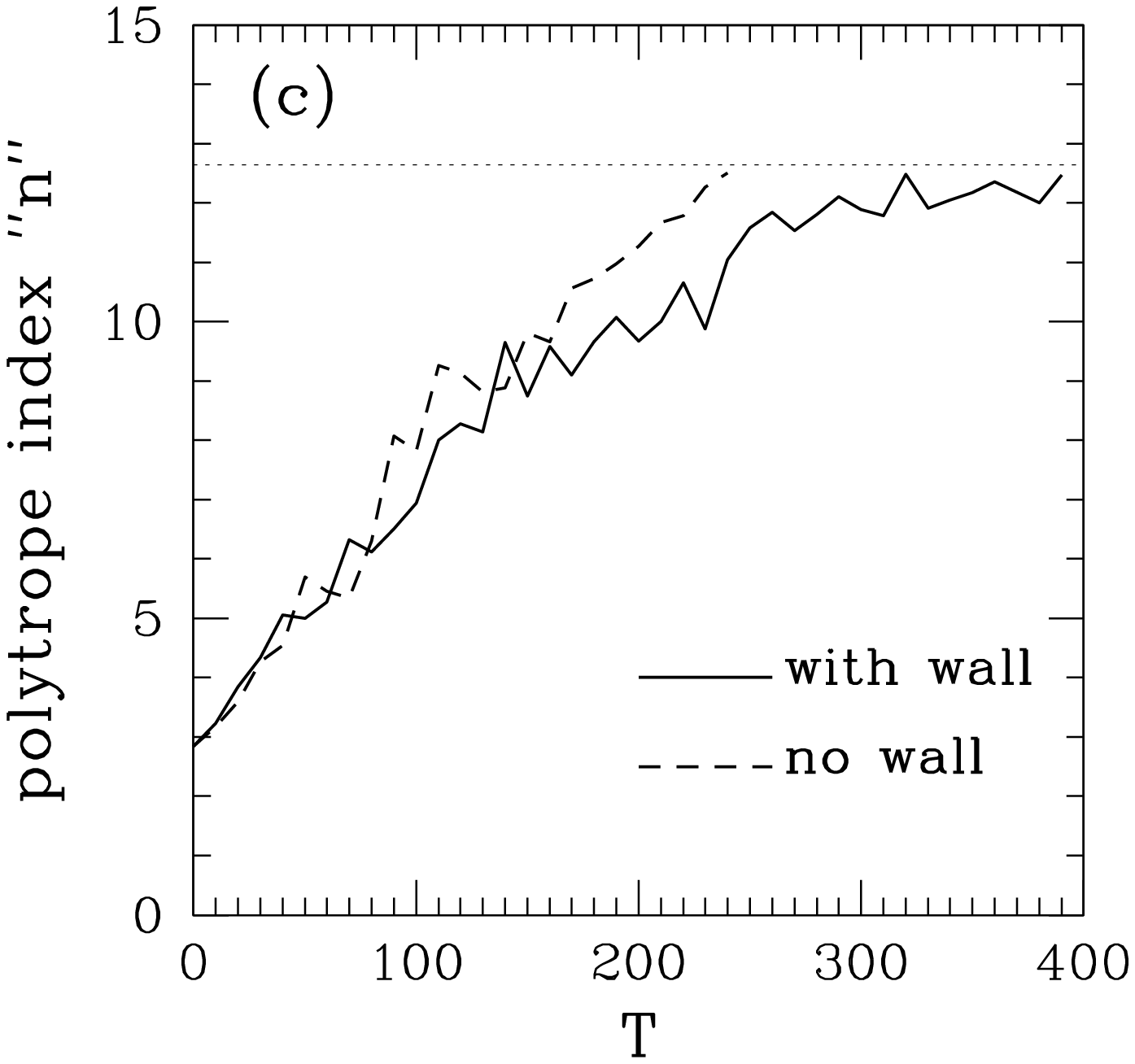}
\caption{\label{fig:run_A} Results from simulation run A. 
(a) Snapshots of density profile $\rho(r)$. 
(b) Snapshots of one-particle distribution function 
$f(\varepsilon)$. 
(c) The time evolution of the polytrope index for run A 
with and without the boundary of adiabatic wall.}
\end{figure*}

Table \ref{tab:initial_model} summarizes 
the list of the five simulation runs. A more systematic 
study of the systems with several initial conditions is now in progress and the 
details of the results will be reported elsewhere. 
In Table \ref{tab:initial_model}, in 
addition to the stellar polytropic  initial state, we also 
consider the Hernquist model\cite{H1990}, 
which was originally introduced to account 
for the empirical law of observed elliptical galaxies\cite{BT_S}. 
%
%
%
%
%
%
%
%
%
%
%
%
%

{\it Results.---} 
It has been recently discussed in \cite{TS2002a,TS2003b,TS2002c} that  
the thermodynamic structure of the stellar polytropic distribution 
can be consistently characterized by the non-extensive framework of the 
thermostatistics. According to their results, the stellar polytrope confined in an 
adiabatic wall is shown to be thermodynamically stable 
when the polytrope index $n<5$. In other words, if $n>5$, a stable 
equilibrium state ceases to exist for a sufficiently large 
density contrast $D>\Dcrit$,   
where the critical value $\Dcrit$ given by a function of $n$ is determined 
from the second variation of entropy around the extremum state of Tsallis 
entropy, $\delta^2S_q=0$\cite{TS2002a,TS2002c}. 
The dotted line in Fig.\ref{fig:lambda} represents 
the critical value $\Dcrit$ for each polytrope index, which indicates that 
the stellar polytrope at low density contrast $D<\Dcrit$ is 
expected to remain stable. Apart from the BG state, 
one might expect that the stellar polytrope acts as a thermal 
equilibrium. 
\begin{figure*}[htb]
\includegraphics[height=4.4cm]{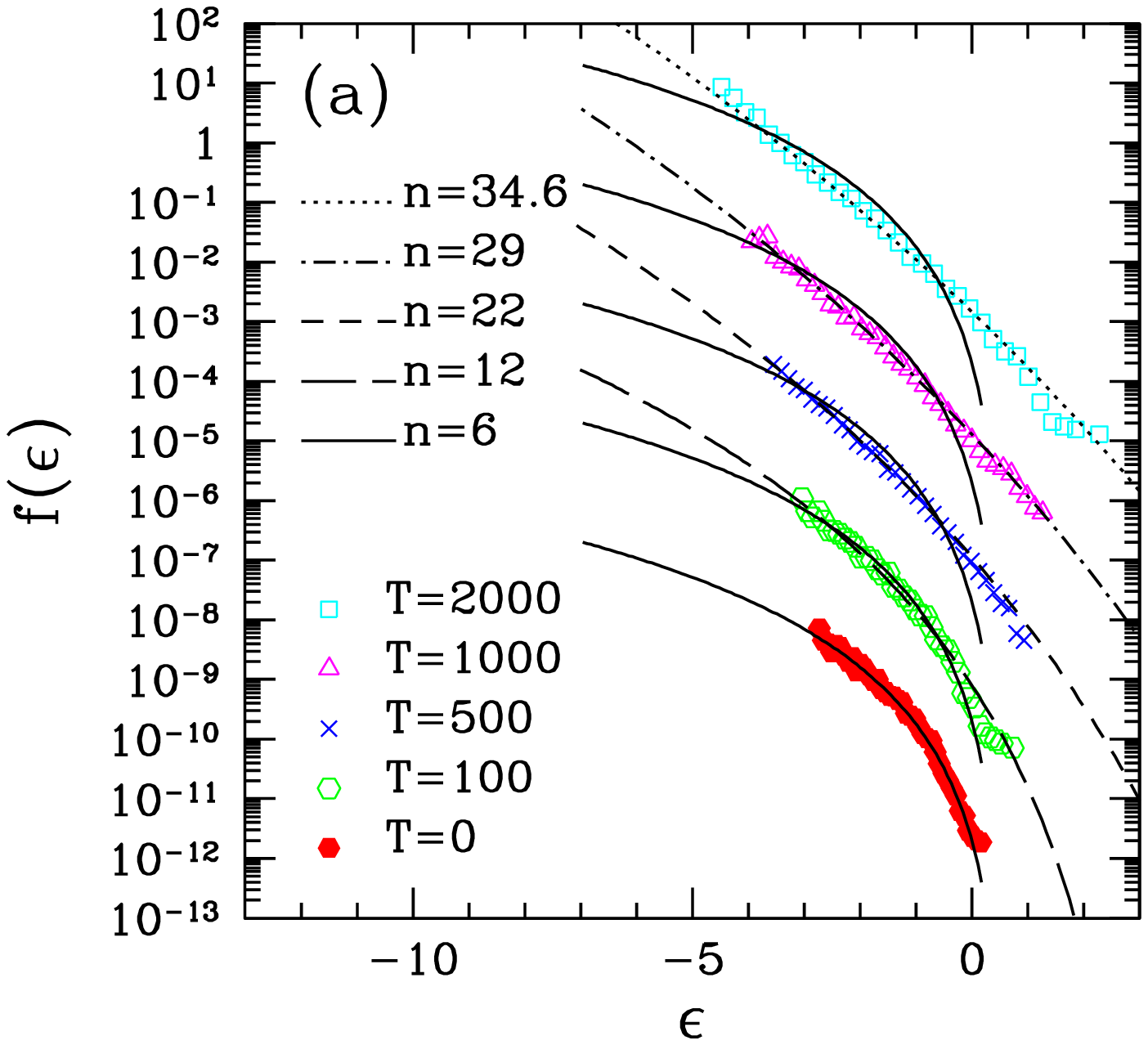}
\hspace*{0.4cm}
\includegraphics[height=4.4cm]{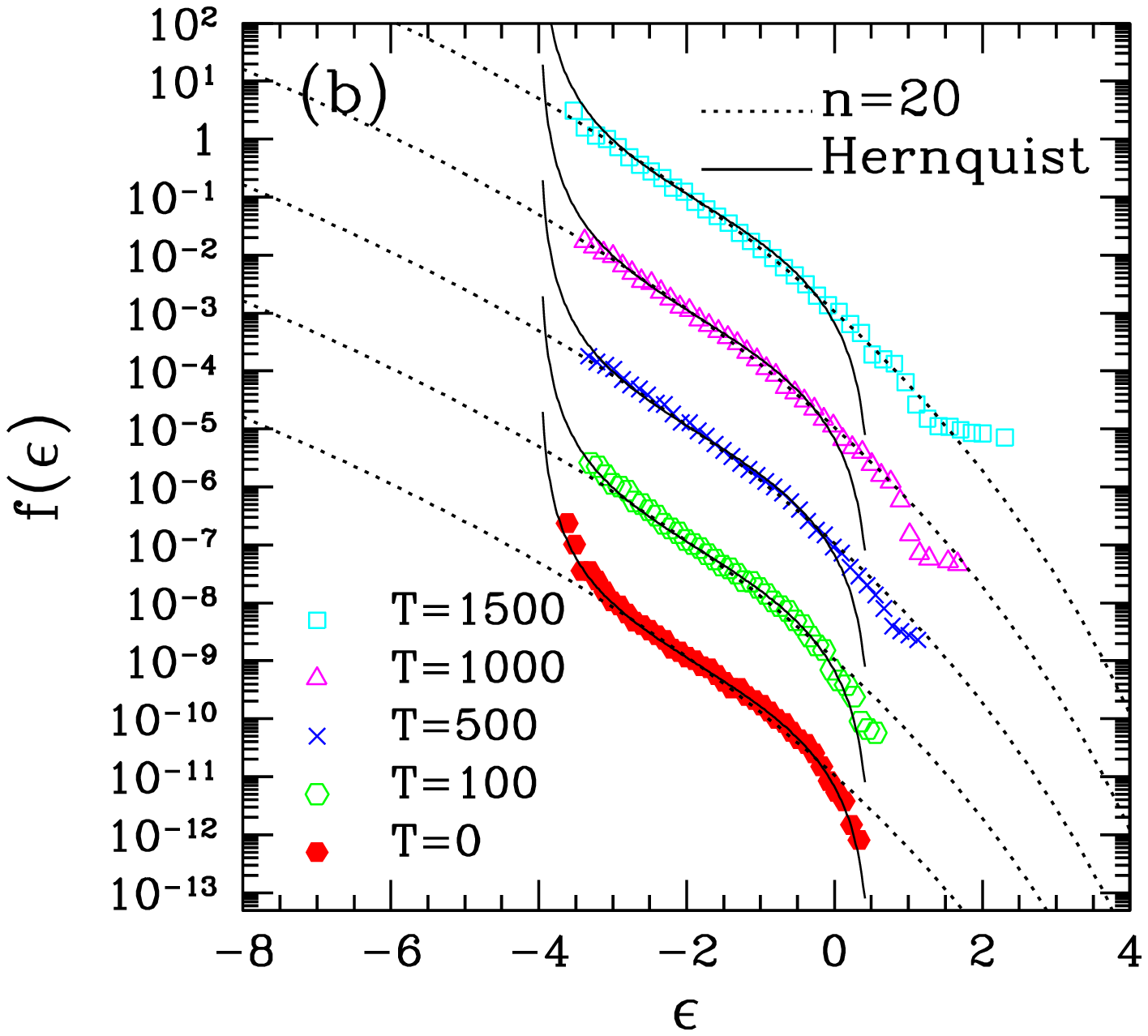}
\hspace*{0.4cm}
\includegraphics[height=4.4cm]{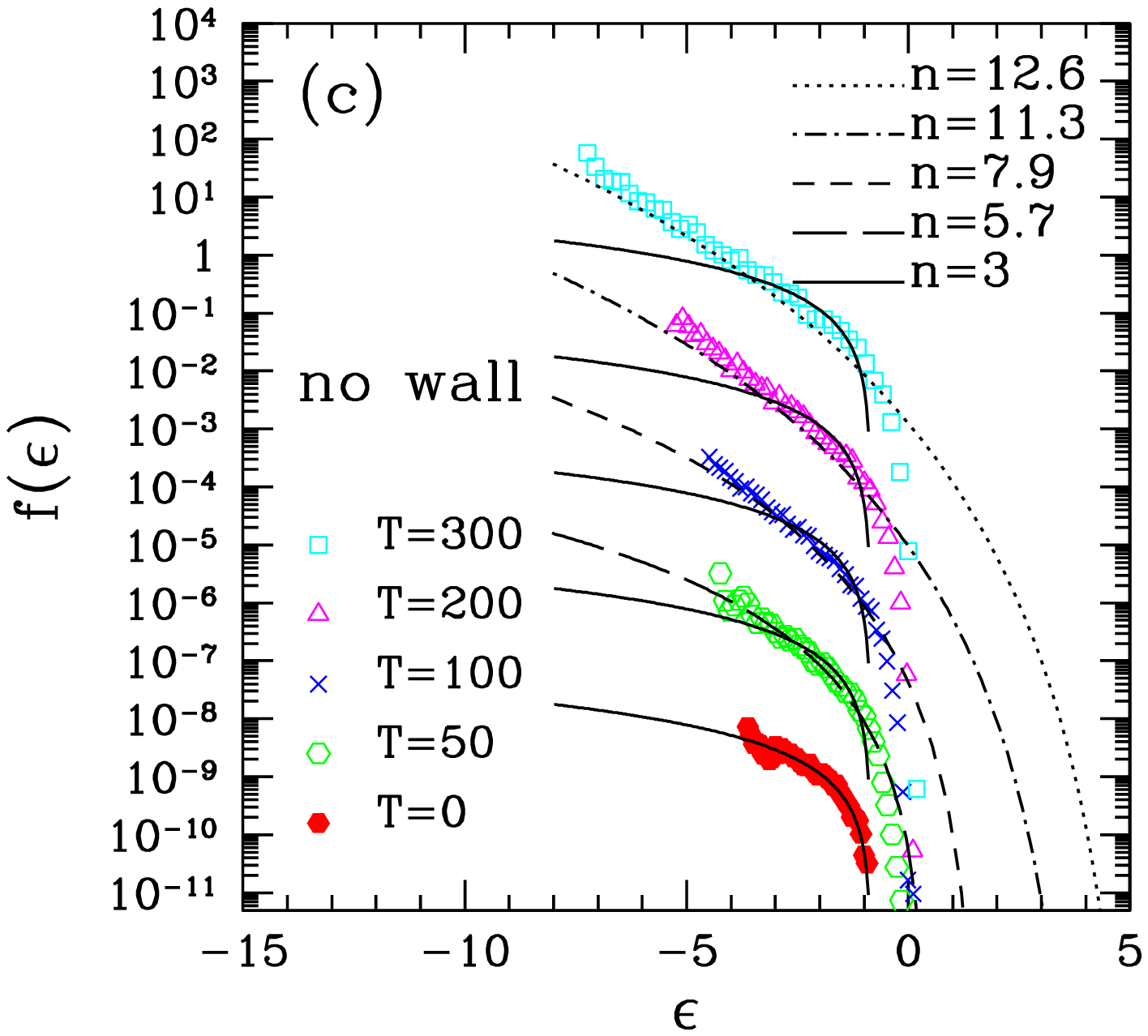}
\caption{\label{fig:fe_others} Evolution of one-particle distribution function in 
other models. (a) Run B1. (b) Run C1. (c) Run A 
removing the adiabatic wall.}
\end{figure*}

Of course, this naive expectation is not correct at all. 
Indeed, the numerical simulations reveal that 
the stellar polytropic distribution gradually changes with time,  
on the timescale of two-body relaxation. 
Further, it seems that the gravothermal instability appears at relatively 
smaller values of $D$ than the predicted one, $D_{\rm crit}$, which might 
be partially ascribed to the non-stationarity of the background stellar 
polytrope. Physically, the core-collapse phenomenon due to the gravothermal 
catastrophe follows from the decoupling of the relaxation timescales 
between the central and the outer parts, whose behavior sensitively depends 
on the physical property of heat transport\cite{MH1991}. In a rigorous 
sense, the thermodynamic prediction 
might lose the physical relevance, however, focusing on the evolutionary 
sequence, we found that the transient state starting 
from the initial stellar polytrope can be remarkably characterized by 
a sequence of stellar polytropes (run A, B1 and B2). 
This is even true in the case starting from the Hernquist model (run C1).

Let us show the representative results taken from run A 
(Fig.\ref{fig:run_A}). 
Fig.\ref{fig:run_A}(a) plots the snapshots of the density profile 
$\rho(r)$, while Fig.\ref{fig:run_A}(b) represents the distribution 
function $f(\varepsilon)$ as a function of the specific energy 
$\varepsilon$. 
Note that just for illustrative purpose, 
each output result is artificially shifting to the two-digits 
below. Only the final output with $T=400$ represents the correct 
scales. In each figure, solid lines mean the initial stellar 
polytrope with $n=3$ and the other lines indicate the fitting 
results to the stellar polytrope by varying the polytrope index 
$n$\cite{notice_3}. 
Note that the number of fitting parameters  
just reduces to one, i.e., the polytrope index, since the total energy 
is well-conserved in the present situation. Fig.\ref{fig:run_A} 
shows that while the system gradually deviates from the initial 
polytropic state, the transient state 
still follows a sequence of stellar polytropes. 
The fitting results are remarkably good until the time exceeds 
$T\simeq400$, corresponding to $15\trelax$. Afterwards, 
the system enters the gravothermally 
unstable regime and finally undergoes the core-collapse.

Now, focus on the evolutionary track in each simulation run 
summarized in the energy-density contrast plane (Fig.\ref{fig:lambda}), 
where the filled circle represents the initial stellar polytrope. 
Interestingly, the density contrast 
of the transient state in run A initially 
decreases, but it eventually turns to increase. 
The turning-point roughly corresponds to the stellar polytrope with index 
$n\sim5-6$. Note, however, that the time evolution of polytrope index itself 
is a monotonically increasing function of time as shown in 
Fig.\ref{fig:run_A}(c), apart from the tiny fluctuations. This is indeed true for 
the other cases, indicating the Boltzmann $H$-theorem 
that any of the self-gravitating systems tends to approach the BG state. 
A typical example is the run C2, which finally reaches the 
stable BG state. However, as already shown in run A, 
all the systems cannot reach the BG state. 
Fig.\ref{fig:lambda} indicates that no BG state is possible 
for a fixed value $\lambda>0.335$\cite{A_LW}, which can be derived from 
the peak value of the trajectory. 
Further, stable stellar polytropes cease to exist at high density contrast 
$D>\Dcrit$. In fact, our simulations starting from the stellar polytropes  
finally underwent core-collapse (runs A, B1 and B2). 
Though it might not be rigorously correct, the predicted value $\Dcrit$ 
provides a crude approximation to the boundary between the stability 
and the instability.

Fig.\ref{fig:fe_others} plots the snapshots of the distribution function taken 
from the other runs. The initial density contrast in run B1 
(Fig.\ref{fig:fe_others}(a)) is relatively low ($D=110$) and thereby the system 
slowly evolves 
following a sequence of stellar polytropes. After $T=2000\sim 74\trelax$, 
the system begins to deviate from the stable equilibrium sequence, 
leading to the core-collapse. Another noticeable case is the run C1 
(Fig.\ref{fig:fe_others}(b)). 
The Hernquist model as initial distribution of run C has cuspy density 
profile, $\rho(r)\propto 1/r/(r+a)^3$, which behaves as $\rho\propto r^{-1}$ 
at the inner part 
\cite{H1990}. The resultant distribution function $f(\varepsilon)$ shows a 
singular behavior at the negative energy region, which cannot be described by the 
power-law distribution. 
Soon after a while, however, the gravothermal expansion\cite{EFM1997}  
takes place and the 
flatter core is eventually formed. Then the system settles into a sequence 
of stellar polytropes and can be approximately described by the stellar 
polytrope with index $n=20$ for a long time. Thus, as long as the 
stellar system confined in an adiabatic wall is concerned,  
the stellar polytrope can be regarded as a quasi-attractor and a 
quasi-equilibrium state.

Of course, these remarkable features could be an outcome in a 
very idealized situation and one suspects that quasi-equilibrium state of 
stellar polytrope 
cannot hold if we remove the boundary of the adiabatic wall. 
As a demonstration, Fig.\ref{fig:fe_others}(c) plots the 
results removing the boundary, in which the initial state is the same 
distribution as in run A. As is expected, the high-energy particles freely 
escape outwards from the central region and the resultant distribution function 
$f(\varepsilon)$ sharply falls off at the energy region $\varepsilon\sim0$, 
indicating that the density contrast $D$ becomes effectively large.   
Thus, compared to the system confined in the wall, the removal of the boundary 
makes the stellar system unstable and the core-collapse takes place 
earlier. 
Nevertheless, focusing on the inner part of the denser region, the 
evolution of the core is not significantly affected by the escape particles at 
the outer part and can be fitted by a sequence of stellar polytropes  
(see also the dashed line in Fig.\ref{fig:run_A}(c)). 
The successful fit to the density profile $\rho(r)$ 
almost remains the same. Hence, the stellar polytrope characterized by 
the Tsallis entropy can be even realized in a realistic situation removing 
the boundary of the adiabatic wall. 
%
%
%
%
%
%
%
%
%
%
%
%
%
%

{\it Summary \& Discussions.---} 
We have performed a set of 
numerical simulation of long-term stellar dynamical evolution 
away from the BG state and found that 
the transient state of the system confined in an adiabatic wall 
can be remarkably fitted by a sequence of stellar 
polytropes. This is even true in the case removing the outer boundary. 
Therefore, the stellar polytropic distribution can be a quasi-attractor 
and a quasi-equilibrium state of a self-gravitating system.

Alternative characterization of the transients 
away from the BG state might be possible besides  
the $q$-exponential distribution of stellar polytropes. 
For an empirical characterization of observed structure, 
the one-parameter family of truncated exponential distributions, 
so-called King model has been used in 
the literature\cite{BT_S,MH,K1966}. Also, the sequence of 
King model has been found to characterize 
the evolutionary sequence of density profile for isolated 
stellar systems without boundary\cite{C1980}. 
We have also tried to fit the simulation data to the 
King model. Similarly to the stellar polytrope, 
the King model accurately describes the simulated density 
profile $\rho(r)$ confined in an adiabatic 
wall, however, it fails to match the simulated distribution function 
$f(\varepsilon)$, especially at the cutoff energy scales. 
Therefore, from the quantitative description of 
the entire phase-space structure, 
the power-law distribution of the stellar polytropes can be a better 
characterization of the quasi-equilibrium state and this could yield an 
interesting explanation of the origin of the empirical King model.

%
%
%
%
%
%
%
%
%
%
%
%
%
%
%
%
%

We are grateful to T. Fukushige for providing us the GRAPE-6 code and 
for his constant supports and helpful comments. 
We also thank to J. Makino for fruitful discussion 
especially on the applicability of equilibrium sequences of stellar 
polytrope. Numerical computations were carried out at ADAC
(the Astronomical Data Analysis Center) 
of the National Astronomical Observatory of Japan. This research was 
supported in part by the grant-in-aid from Japan Society of Promotion of 
Science (No.1470157).
%
%
%
%
%
%
%
%
%
%
%
%
%
%
%
%
%
%
%
%


\begin{thebibliography}{}
\bibitem{notice_3} In fitting the simulation data to the stellar polytrope, 
	we first quantify the radial density profile $\rho(r)$ from each 
	snapshot data. Selecting the $100$ points from it 
	at regular intervals in logarithmic scale of radius $r$, 
	the results are then compared with 
	the Emden solutions, fixing the energy $\lambda$  and varying the 
	polytrope index $n$.
\end{thebibliography}
\end{document}